\documentclass[aps,preprint,tightenlines,showpacs,nofootinbib]{revtex4}
\usepackage{epsfig}
\usepackage{multirow}
\usepackage{amsmath,stackrel}
\usepackage{amssymb}
\usepackage{graphicx}
\usepackage{soul}

\begin{document} 

\title{Constituent-quark model description of triply heavy-baryon nonperturbative lattice QCD data}
\author{J.~Vijande}
\affiliation{Departamento de F\'{\i}sica At\'{o}mica, Molecular y Nuclear, Universidad de Valencia (UV)
and IFIC (UV-CSIC), E-46100 Valencia, Spain}
\author{A.~Valcarce}
\affiliation{Departamento de F\'\i sica Fundamental and IUFFyM, Universidad de Salamanca, E-37008
Salamanca, Spain}
\author{H.~Garcilazo}
\affiliation{Escuela Superior de F\'\i sica y Matem\'aticas,
Instituto Polit\'ecnico Nacional, Edificio 9,
07738 M\'exico D.F., Mexico}
\date{\emph{Version of }\today}

\begin{abstract}
This paper provides results for the spectra of triply
charmed and bottom baryons based on a constituent quark model approach.
We take advantage of the assumption that potential models are expected to describe triply heavy baryons 
to a similar degree of accuracy as the successful results obtained in the charmonium and bottomonium sectors. 
The high precision calculation
of the ground state and positive and negative parity excited states recently reported by nonperturbative lattice QCD
provides us with a unique opportunity to confront model predictions
with data. This comparison may also help to build a bridge between two difficult to reconcile lattice QCD results, 
namely, the lattice SU(3) QCD static three-quark potential and the recent results of
nonperturbative lattice QCD for the triply heavy-baryon spectra.
\end{abstract}
\maketitle

\section{Introduction}
In a previous publication~\cite{Vij15} we have pointed
out that the static three-quark potential with parameters
determined from SU(3) lattice QCD~\cite{Tak02} does not reproduce
the triply heavy-baryon $bbb$ and $ccc$ spectra measured also in 
lattice QCD~\cite{Mei10,Mei12,Pad13}. We argued several possible 
reasons for such disagreement. In this work we aim to analyze
whether the triply heavy-baryon spectra recently calculated
by means of nonperturbative lattice QCD techniques can be 
understood within a constituent quark-model framework by 
means of simple Cornell-like potentials as it is the case
of charmonium and bottomonium spectra~\cite{Qui79,Eic80,Eic08}.
This calculation may help to connect the description of the
heavy-baryon spectra obtained by nonperturbative lattice QCD techniques
and the static potentials derived within SU(3) lattice QCD.

Theoretically, one would expect that potential models would be able
to describe triply heavy baryons to a similar degree of precision as 
their success in charmonium and bottomonium. As noticed by Bjorken 
some time ago~\cite{Bjo85}, bound states of three-heavy quarks reveal 
a pure baryonic spectrum without light-quark complications. In the 
same way the $Q\bar Q$ interactions are examined in heavy mesons, 
the study of triply heavy baryons will probe the $QQ$ interactions in the 
heavy quark sector. However, no experimental results are available so far 
for triply heavy baryons (see Ref.~\cite{Che11} for a recent calculation 
of production cross section at the LHC), and thus, the predictions of 
their properties cannot yet be compared to the real world. 
The recent precise calculation of the ground and excited states of
triply bottom baryons~\cite{Mei10,Mei12}, together with the ground and excited states of 
triply charm baryons~\cite{Pad13}, provides us with 
a unique opportunity to test phenomenological 
quark models for baryons in the energy regime in which the description 
using potential models is expected to work best, the heavy-quark sector.
Hence, the quark-model dependent calculations could be 
tested by comparing them to nonperturbative first-principles calculations 
in lattice QCD of the $bbb$ and $ccc$ systems.

The paper is organized as follows. In the next section we will briefly
described the Cornell potential used for the description of the heavy-meson
sector. We will try to connect the well established parameters used in the 
literature for the heavy-meson sector into the relatively unknown 
heavy-baryon sector. In Sec.~\ref{Comp} we will present our results. We will firstly analyze 
the pattern of the nonperturbative lattice QCD results looking for the general structure of the 
potential. We will pursue different parametrizations in a trial
to get a unified description of the $bbb$ and $ccc$ spectra. The results will 
be derived by different numerical techniques previously tested by our
group: generalized Gaussians variational approaches, hyperspherical
harmonics and Faddeev equations~\cite{Gar07,Val08,Vij09}.
Finally, in Sec.~\ref{Res}, we will summarize the 
main conclusions of this study. 

\section{A constituent quark model potential for three-heavy quarks.}

Since the early days of QCD the interaction among heavy quarks has been explored
as an important tool to learn about the behavior of QCD at low energies. 
At the end of 1974, when the new particles seen
at Brookhaven and SLAC were identified as $c\bar c$ bound states, explicit
models were proposed to calculate the spectrum and the radiative transitions 
(see Ref.~\cite{Ric12} for a recent pedagogical review).
The potential proposed in~\cite{Eic75} is known as the funnel or Cornell potential
and it reads,
\begin{equation}
V^{Q\bar Q}_{ij}(r) = -\frac{a}{|\vec{r}_i - \vec{r}_j|} + b\, |\vec{r}_i - \vec{r}_j| + c \, .
\end{equation}
Solving the two-body problem, one can tune the parameters to reproduce the 
low levels of charmonium. This was done by several groups in the 70's and 
the authors were able to predict the missing states. The game became more 
challenging when the first bottomonium levels were found, trying to reproduce
simultaneously the $c\bar c$ and $b\bar b$ spectra. Indeed, even so the
interquark potential was not derived from QCD in early quarkonium phenomenology,
it was assumed to be universal, or flavor independent. In QCD the gluons
couple to the color, hence it is reasonable to assume that the potential is
flavor independent.

The $Q\bar Q$ potential has also been extensively studied by lattice gauge 
theories~\cite{Bal01}, being nowadays a very well-known quantity resembling the 
structure derived from the heavy-meson spectra. Thus, the typical 
shape of the color-singlet $Q \bar Q$ potential is characterized by
a short-range Coulomb behavior and a long-range linear rise, that well represents the 
double nature of QCD as an asymptotically free and infrared confined theory. 
The excitation spectrum of the gluon field around a static quark-antiquark pair 
has also been explored by lattice calculations~\cite{Jug03}.
On the large length scale the spectrum agrees with 
that expected for string-like excitations while
in the short range it shows a Coulomb-like behavior 
as it was first noted within the context of the static 
bag picture of gluon excitations~\cite{Has80}. 

The $3Q$ potential should be the analog of the famous Cornell potential
for quarkonium. The short-distance behavior of $V^{3Q}(r)$ is expected to be 
described by the two-body Coulomb potential as the one-gluon exchange result 
in perturbative QCD. It should be extended for the baryon case, with a 
factor $1/2$ in front of its strength due to color factors~\cite{Ric12}. 
As for the $Q\bar Q$ case, 
the characteristic signature of the long-range non-Abelian dynamics is 
believed to be a linear rising of the static interaction. Moreover, 
the general expectation for the baryonic case is that, at least 
classically, the strings meet at the 
so-called Fermat (or Torricelli) point, which has minimum
distance from the three sources ($Y$-shape configuration)~\cite{Tak04,Bor04}. 
The confining short-range $3Q$ potential could be also 
described as the sum of two-body potentials 
($\Delta-$shape or linear configuration)~\cite{Tak04,Bor04,Cor04,Ale03}. 
We have demonstrated in Ref.~\cite{Vij15} the equivalence of 
both prescriptions for the case of triply heavy baryons (see 
Table II of that reference) and for different values
of the heavy-quark mass.
Thus, a minimal model to study $3Q$ systems may come given by,
\begin{equation}
V^{3Q}(r) = - A \sum_{i < j}\frac{1}{|\vec{r}_i - \vec{r}_j|} \, + \, B 
\sum_{i < j}|\vec{r}_i - \vec{r}_j| \, + \, C \,  .
\label{Pot3Q}
\end{equation}
The value of the $Q\bar Q$ confinement strength $b$ is usually fixed to reproduce that obtained
from the linear Regge trajectories of the pseudoscalar $\pi$ and $K$ mesons, $\sqrt{\sigma}=$
(429$\pm$2) MeV~\cite{Bal01}. In the case of baryons, the linear 
string tension $B$ is considered to be of the order of
a factor $1/2$ of the $Q\bar Q$ case. The reduction factor in the string tension can be naturally understood as a geometrical
factor rather than a color factor, due to the ratio between the minimal distance joining three heavy quarks and the 
perimeter length of a $3Q$ triangle, suggesting 
$B= \left(0.50 \sim 0.58 \right) b$~\cite{Tak02}.
For the particular case of quarks in an equilateral triangle 
$B= \frac{1}{\sqrt{3}} \, b = 0.58 \, b $~\cite{Bor04}.
When the linear ansatz is adopted for the
two-body potential, still the same relation holds for the strength of the Coulomb potential $A\simeq \frac{1}{2} a$, due to color factors. 
The $\Delta$ ansatz (linear potential) has been widely adopted in the nonrelativistic quark model because of its 
simplicity~\cite{Gar07,Val08,Isg78,Oka81,Sil96,Kle10,Cre13}

\section{Results and discussion}
\label{Comp}

To check whether the Cornell-like potential of Eq.~(\ref{Pot3Q}) reproduces the
$bbb$ and $ccc$ baryon spectra measured in lattice QCD~\cite{Mei10,Mei12,Pad13} we will make
use of three different numerical methods: the generalized Gaussians variational approach~\cite{Vij09}, 
hyperspherical harmonics~\cite{Val08} and Faddeev equations~\cite{Gar07}. The three methods have
been used and the difference in results is negligible.
In all cases we solve the nonrelativistic Schr\"odinger equation
\begin{equation}
\left\{H_0 + V^{3Q}(r)\right\}\Psi(\vec r) = E \Psi({\vec r}) \, ,\nonumber
\end{equation}
where $H_0$ is the free part of three-heavy quarks without center-of-mass-motion
\begin{equation}
H_0=\sum_{i=1}^3 \left( M_{Q} + \frac{\vec{p}_i^{\,2}}{2M_{Q}} \right) - T_{CM}\, \nonumber
\end{equation}
and $M_{Q}$ is the mass of the heavy quark. The mass of the heavy baryon will be finally 
given by $M_{3Q}=3M_Q + E$.

It has been demonstrated in Ref.~\cite{Vij15} that the results of nonperturbative 
lattice QCD for the $bbb$ and $ccc$ systems are not reproduced by the static three-quark
potential derived in SU(3) lattice QCD~\cite{Tak02}. 
We had hypothesized on several possible reasons for this disagreement. Among them, we
suggested the possibility that the parameters
obtained in quenched QCD in Ref.~\cite{Tak02} might
be different from those in 2+1 flavor QCD employed in Refs.~\cite{Mei10,Mei12,Pad13}.
To check this possibility parameters in 2+1 flavor QCD has to be obtained, unfortunately such calculations do not exist so far. 
We also wondered about the 
possibility that the fitting form Eq.~(\ref{Pot3Q}) might not be appropriate to describe
the static three-quark potential in lattice QCD or that
the quark model description with "3-quark potential" might be
inappropriate for $bbb$ and $ccc$ systems.
We will try to mark out these possibilities by studying the nonperturbative lattice QCD data
in terms of the suggested Cornell-like potential of Eq.~(\ref{Pot3Q}) with parameters adjusted 
to the data, looking also for the possible need of higher order terms.

In Ref.~\cite{Vij15} it has been noticed how $bbb$ nonperturbative lattice QCD calculations point to a $3Q$ potential
given by a mixture of a linear confinement and a Coulomb interaction.
The comparison performed of the $bbb$ and $ccc$ nonperturbative 
lattice QCD results and calculations based on a mixture of a $\Delta$-shape confinement and a Coulomb 
interaction using the parameters reported in Ref.~\cite{Tak02} ($A=$ 0.1410 and 
$B=$ 0.0925 GeV$^2$) showed a large difference in the excited states, predicting a 
small splitting between positive and negative parity excited states
and also a small excitation energy for the positive parity states. These 
results may point to a lack of strength either in the confining or in the 
Coulomb potential.
\begin{figure}[t]
\vspace*{-7cm}
\hspace*{-1cm}\mbox{\epsfxsize=180mm\epsffile{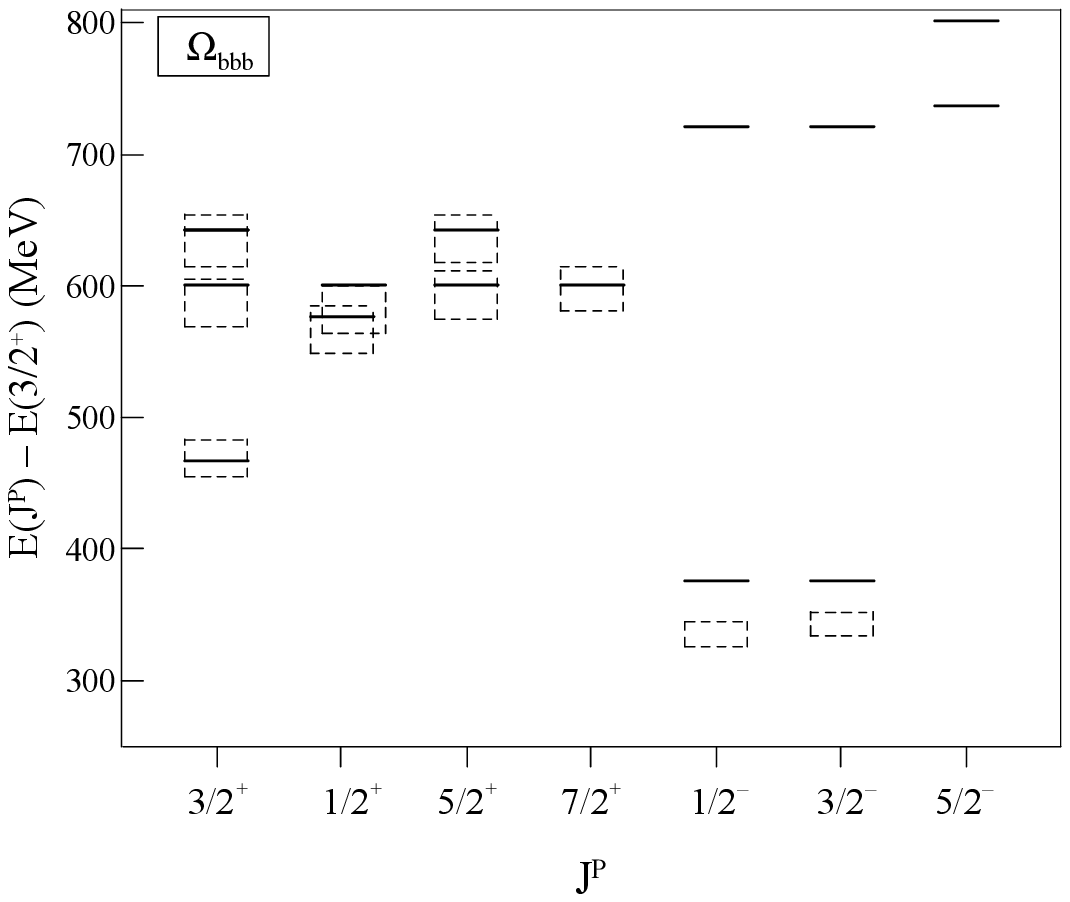}}\\
\vspace*{-17.cm}
\hspace*{-1cm}\mbox{\epsfxsize=180mm\epsffile{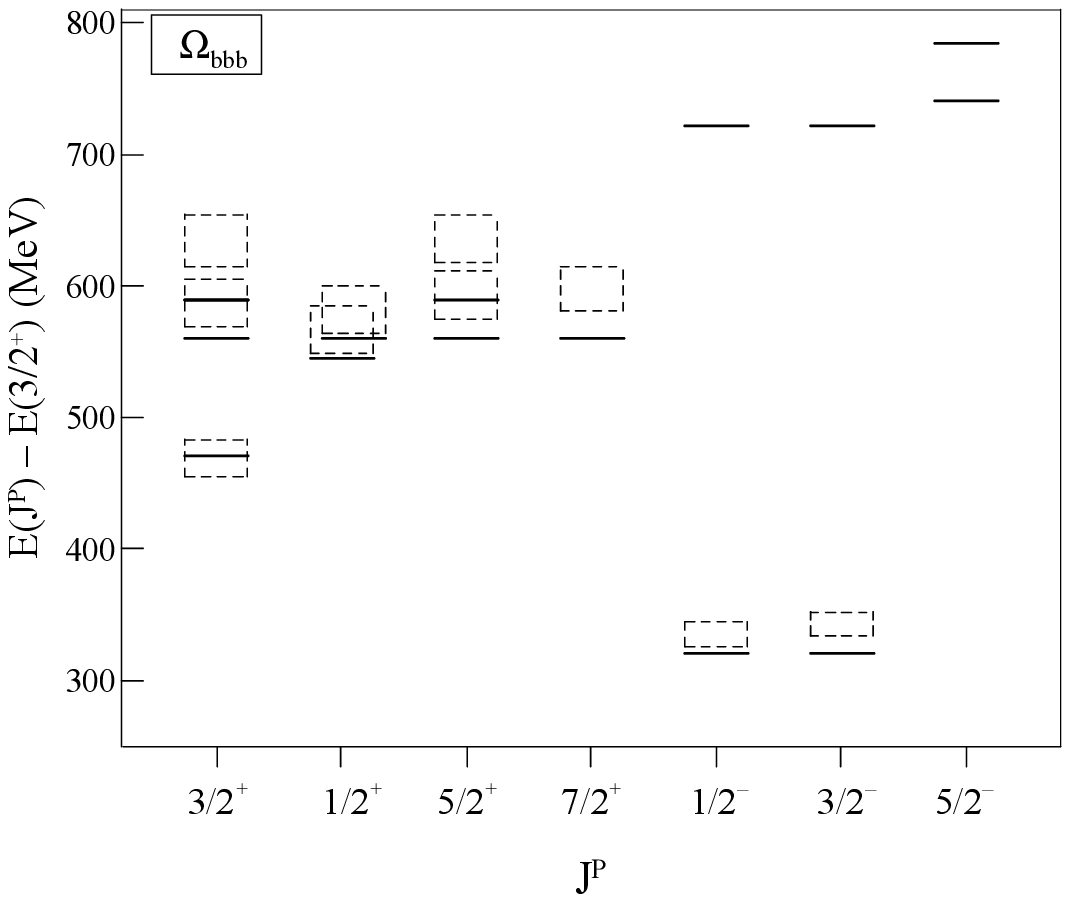}}
\vspace*{-12.cm}
\caption{$bbb$ excited state spectra, solid lines, for the potential of Eq.~(\ref{Pot3Q}) with
$A=$ 0.2787 and $B=$ 0.0925 GeV$^2$
(upper panel) or $B=$ 0.1517 GeV$^2$ and  $A=$ 0.1410 (lower panel).
See text for details.
The boxes stand for the nonperturbative lattice QCD results of Ref.~\cite{Mei12}.}
\label{fig1}
\end{figure}
\begin{figure}[t]
\vspace*{-7cm}
\hspace*{-1cm}\mbox{\epsfxsize=180mm\epsffile{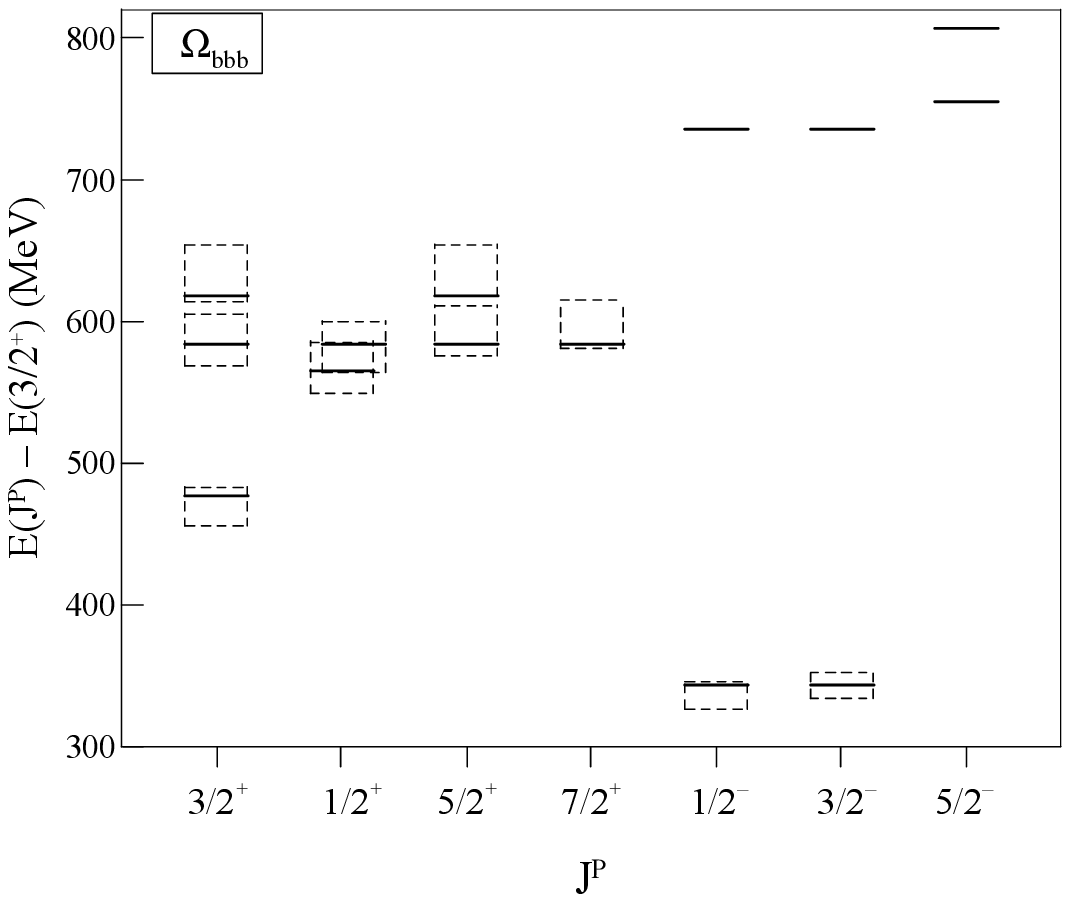}}\\
\vspace*{-17.cm}
\hspace*{-1cm}\mbox{\epsfxsize=180mm\epsffile{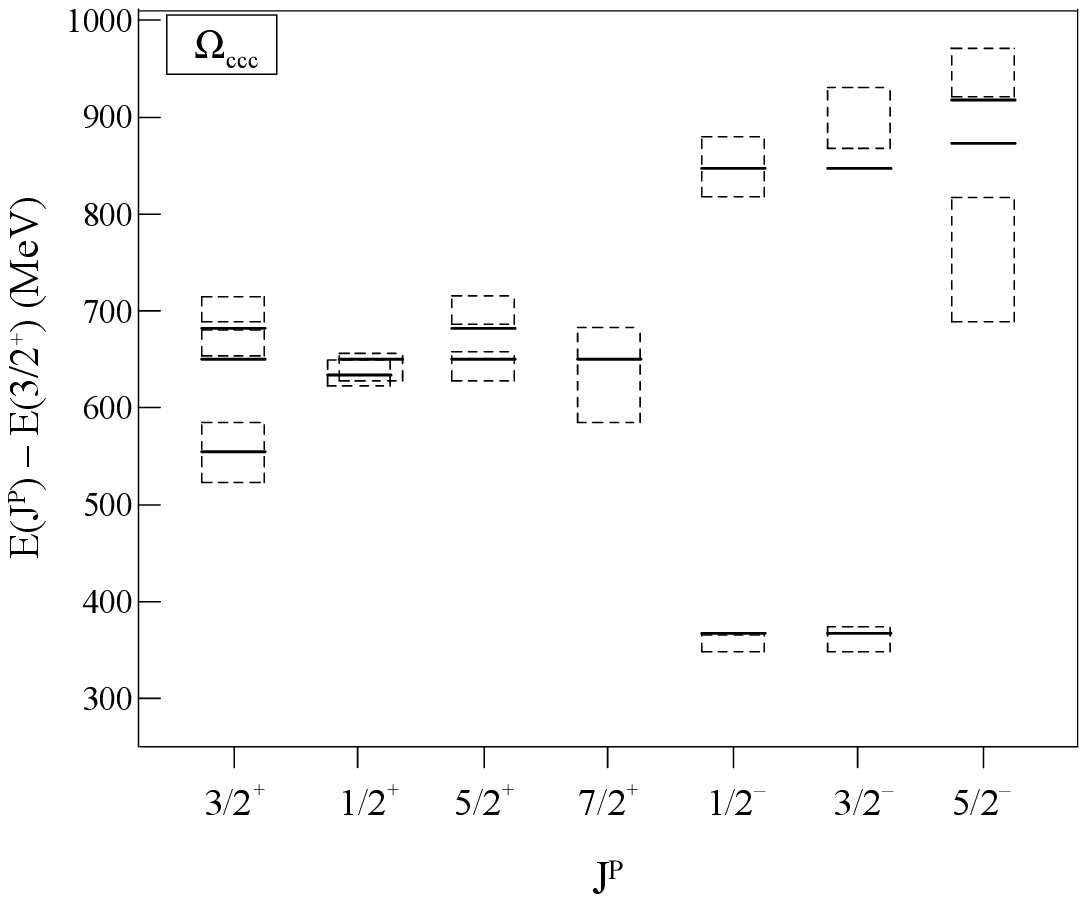}}
\vspace*{-12.cm}
\caption{$bbb$ (upper panel) and $ccc$ (lower panel) excited state spectra, solid lines, 
for the  potential of Eq.~(\ref{Pot3Q}) with the parameters $A=$ 0.1875 and 
$B=$ 0.1374 GeV$^2$. The boxes stand for the nonperturbative lattice QCD results 
of Ref.~\cite{Mei12} for the $bbb$ system and Refs.~\cite{Pad13,Mat14} for the $ccc$ system.}
\label{fig2}
\end{figure}

We have analyzed which of the two terms, linear confinement or Coulomb, would 
play the relevant role by keeping the strength
of one of them as predicted by SU(3) lattice QCD and increasing the other 
to pursue a spectrum close to the nonperturbative lattice QCD results
of Refs.~\cite{Mei10,Mei12,Pad13}. We show in Fig.~\ref{fig1} results 
for the $bbb$ system. In the upper panel we have kept constant the strength of the
linear confining interaction, $B=$ 0.0925 GeV$^2$, and we have augmented
the strength of the Coulomb potential up to $A=$ 0.2787, a value rather close
to the $Q\bar Q$ Coulomb strength~\cite{Tak02}. Alternatively, in the lower 
panel we have kept constant the strength of the
Coulomb potential $A=$ 0.1410, and we have increased the strength of the linear 
confining interaction up to $B=$ 0.1517 GeV$^2$, which is also rather close 
to the $Q\bar Q$ strength~\cite{Tak02}. One can easily identify the peculiarities 
of three identical heavy-quark baryons.
By increasing the Coulomb potential, upper panel, the mass of all excitations 
is increased but the mass difference between the positive and negative parity excited states, 
$E_2(3/2^+) - E_1(1/2^-)$, is smaller than predicted by nonperturbative lattice QCD. This 
mass difference could be enlarged by increasing the strength of the linear
confining potential as we have done in the lower panel, but without lowering too much the mass
of the negative parity states, $E_1(1/2^-)$ and $E_1(3/2^-)$. It seems therefore that a moderate
combined increase of the strength of the two terms, Coulomb and confinement, may allow to get
a better fit to the nonperturbative lattice QCD data.

Thus, we have found a nice fit of the nonperturbative lattice QCD results
with the potential of Eq.~(\ref{Pot3Q}) and the
parameters $A=$ 0.1875 and $B=$ 0.1374 GeV$^2$. 
The results are shown in Fig.~\ref{fig2} for the $bbb$ and $ccc$ systems. 
At this point one should bear in mind that the description of the charmonium 
and bottomonium experimental data with a Cornell-like
potential needs of a large value of the $Q\bar Q$ Coulomb strength, 
$a \simeq 0.51-0.52$~\cite{Qui79,Eic80,Eic08}, as compared to 
that obtained in SU(3) lattice QCD~\cite{Tak02}.
Using the phenomenological values reproducing
the experimental data of the heavy quark meson spectra and the nonperturbative lattice QCD
results for the triply heavy baryon spectra,
one would conclude that $A/a < 1/2$,
slightly different from $1/2$ as the one-gluon exchange result. 
This conclusion could have been already anticipated from models designed to describe the light and
strange baryon spectra and later extrapolated to triply-heavy baryons. 
The $AL1$ model of Ref.~\cite{Sil96} made use of $A=$ 0.2534 and $B=$ 0.0827 GeV$^2$.
We show in Table~\ref{tab3} the triply heavy baryon spectra obtained with this potential model as compared to the
recent nonperturbative lattice QCD results. We observe how for the $bbb$ system,
it drives to a too high negative parity states together with a small splitting between
positive and negative parity states (a similar situation to the upper panel of
Fig.~\ref{fig1}) as a consequence of the large strength of the Coulomb potential.
Besides, for the $ccc$ system this model drives to rather
low excited states as a consequence of the small strength of confinement. 

\begin{table}[t]
\caption{Excited states, in MeV, of $ccc$ and $bbb$ systems from nonperturbative lattice QCD,
Refs.~\cite{Pad13,Mat14} and Ref.~\cite{Mei12} and results for the $AL1$
potential model of Ref.~\cite{Sil96}.}
\begin{center}
\begin{tabular}{|c|p{0.5cm}cp{0.5cm}cp{0.5cm}|p{0.5cm}cp{0.5cm}cp{0.5cm}|}
\hline
             & & \multicolumn{3}{c}{$ccc$}   && & \multicolumn{4}{c|}{$bbb$} \\
             & & Ref.~\cite{Sil96} & & Refs.~\cite{Pad13,Mat14} && & Ref.~\cite{Sil96}  & & Ref.~\cite{Mei12}  & \\ \hline
 $E_2(\frac{3}{2}^+)-E_1(\frac{3}{2}^+)$     & &   472             & &  554$\pm$31              && & 451                & &  469$\pm$14        & \\
 $E_3(\frac{3}{2}^+)-E_1(\frac{3}{2}^+)$     & &   567             & &  667$\pm$13              && & 581                & &  587$\pm$18        & \\
 $E_1(\frac{1}{2}^+)-E_1(\frac{3}{2}^+)$     & &   542             & &  636$\pm$13              && & 555                & &  567$\pm$18        & \\
 $E_1(\frac{1}{2}^-)-E_1(\frac{3}{2}^+)$     & &   316             & &  357$\pm$9               && & 360                & &  335.5$\pm$9.5     &  \\
\hline
\end{tabular}
\end{center}
\label{tab3}
\end{table}

The description of the $bbb$ system shown in Fig.~\ref{fig2} has been obtained with a reasonable mass for the $b$ quark,
$m_b=4.655$ GeV, however in the case of the $ccc$ system one needs a large unrealistic mass 
for the $c$ quark, $m_c=2.050$ GeV, what would drive to a $ccc$ ground state with a mass around 7 GeV.
This seems to indicate that triply charm baryons present a more involved structure that does
not fit so nicely in a simple pairwise linear plus Coulomb interaction. Such deviation may has partially 
its origin on the fact that the lattice QCD calculation of Ref.~\cite{Pad13} did not address all 
the systematic uncertainties, but only statistical uncertainties are given. 
In fact, a calculation of the charmonium spectrum with the same lattice action and the same 
lattice spacing can be found in Ref.~\cite{Liu12} where one can get
an idea of the typical size of the systematic uncertainties.
On the other hand, potential models probably are also less accurate 
for $ccc$ than for $bbb$ baryons, because the $ccc$ system is more relativistic 
and spin-dependent contributions may start playing a significant role. 
Although of small importance in heavy quark systems for being suppressed
as $M_Q^{-2}$, the spin-spin interaction derived from the OGE 
may help to improve the description of the nonperturbative lattice QCD results.
One may therefore add a spin-spin term to the interacting potential, having 
the final form,
\begin{equation}
V^{3Q}_{SS}(r) = - A\sum_{i < j}\frac{1}{|\vec{r}_i - \vec{r}_j|} \, + \, B 
\sum_{i < j}|\vec{r}_i - \vec{r}_j| \, + 
\frac{A}{M_Q^2}\frac{e^{-r/r_0}}{rr_0^2} \vec\sigma_i\cdot\vec\sigma_j \,  .
\label{Pot3QSS}
\end{equation}
\begin{figure}[t]
\vspace*{-6cm}
\hspace*{-1cm}\mbox{\epsfxsize=180mm\epsffile{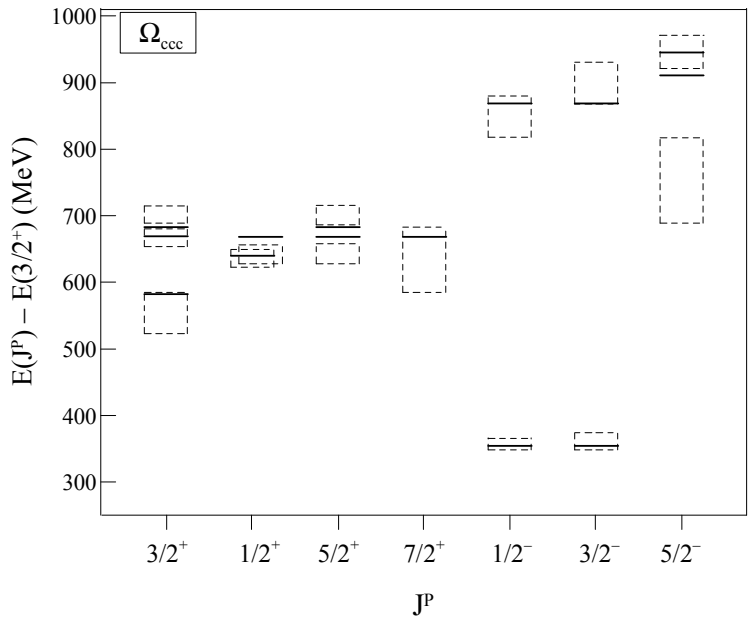}}
\vspace*{-12.cm}
\caption{$ccc$ excited state spectra, solid lines, 
for the potential of Eq.~(\ref{Pot3QSS}) including a spin-spin interaction. 
The boxes stand for the results of Ref.~\cite{Pad13}.}
\label{fig3}
\end{figure}
\begin{figure}[t]
\hspace*{-1cm}\mbox{\epsfxsize=83mm\epsffile{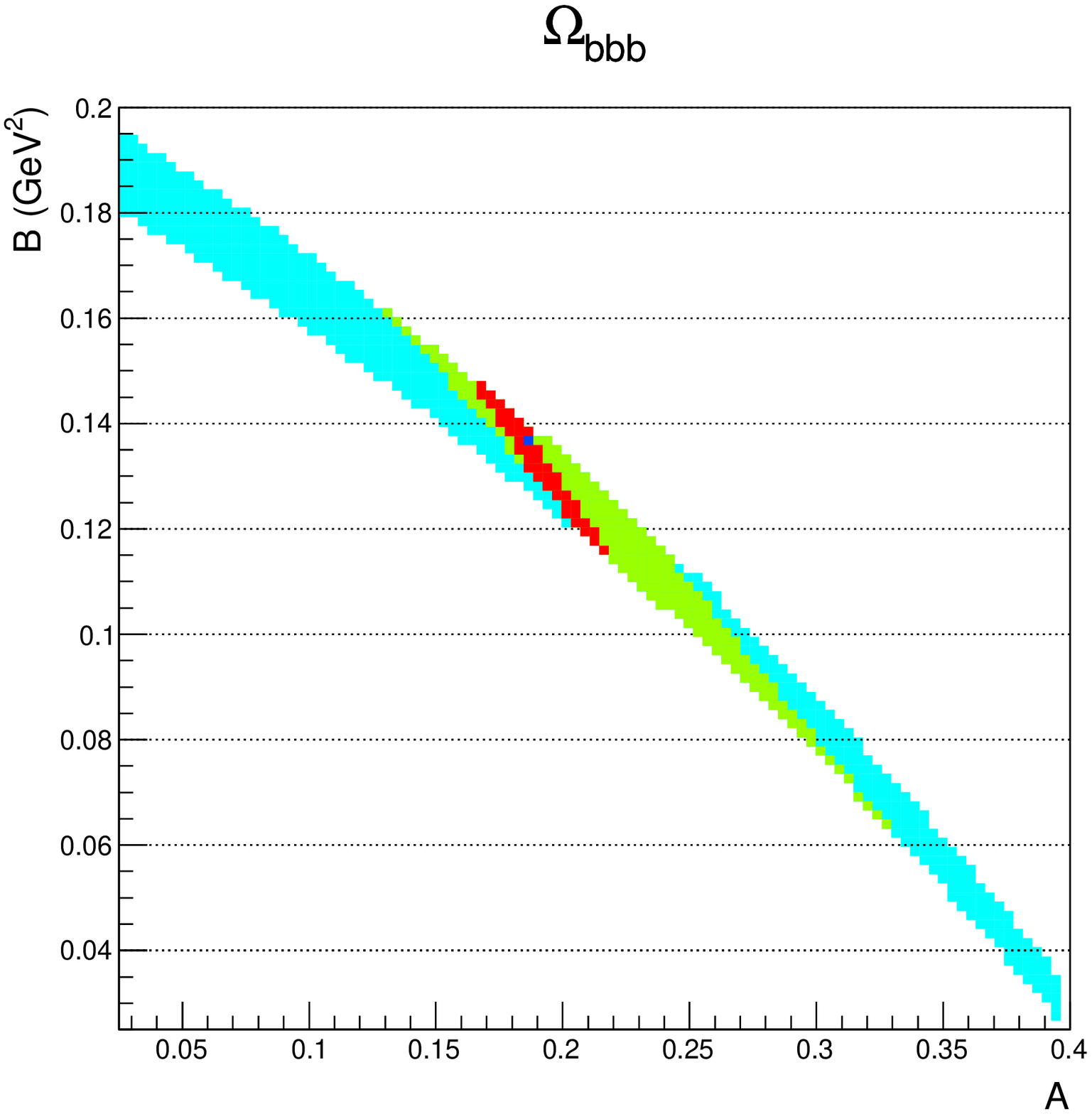}}
\caption{Coulomb and linear confining strengths reproducing selected 
states of the $bbb$ system. The light blue area stands for sets of parameters reproducing the $E_2(3/2^+)$ energy,
the green area reproduces the $E_2(3/2^+)$ and $E_3(3/2^+)$ energies,
the read surface reproduces the $E_2(3/2^+)$, $E_3(3/2^+)$, and 
$E_1(1/2^-)$ energies. The dark blue point corresponds to our best fit.}
\label{fig4}
\end{figure}
We have performed a fit to the $ccc$ data using the same Coulomb and linear confining
strengths used for the $bbb$ system, $A=$ 0.1875 and $B=$ 0.1374 GeV$^2$. 
When the spin-spin term is considered, one can reduce the value of the charm quark mass
preserving a nice agreement with data. We show in Fig.~\ref{fig3} the results
for standard values of quark-potential models, $m_c=1.6$ GeV and $r_0=0.28$ fm~\cite{Sil96}.
As one can see there is a good agreement, sustaining the result of a short-range 
Coulomb potential supplemented by a linear confining interaction. The short-ranged
spin-spin interaction helps to the fine tuning of the negative parity excited 
states (where it is negligible) with respect to the positive parity excited states (where it
is attractive) due to the existence of good {\it diquarks} with spin 0 in the
wave function.
\begin{figure}[t!]
\vspace*{1cm}
\hspace*{-1cm}\mbox{\epsfxsize=85mm\epsffile{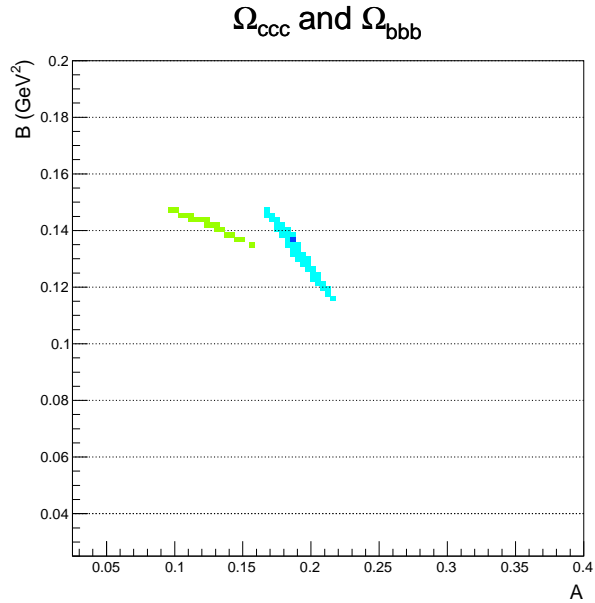}}
\caption{The light blue area stands for Coulomb and linear confining strengths reproducing the 
$E_2(3/2^+)$, $E_3(3/2^+)$, and $E_1(1/2^-)$ energies of the $bbb$ system.
The green surface represents the same for the $ccc$ system. The dark blue point corresponds to our best
fit.}
\label{fig5}
\end{figure}
\begin{figure}[t!]
\hspace*{-1cm}\mbox{\epsfxsize=85mm\epsffile{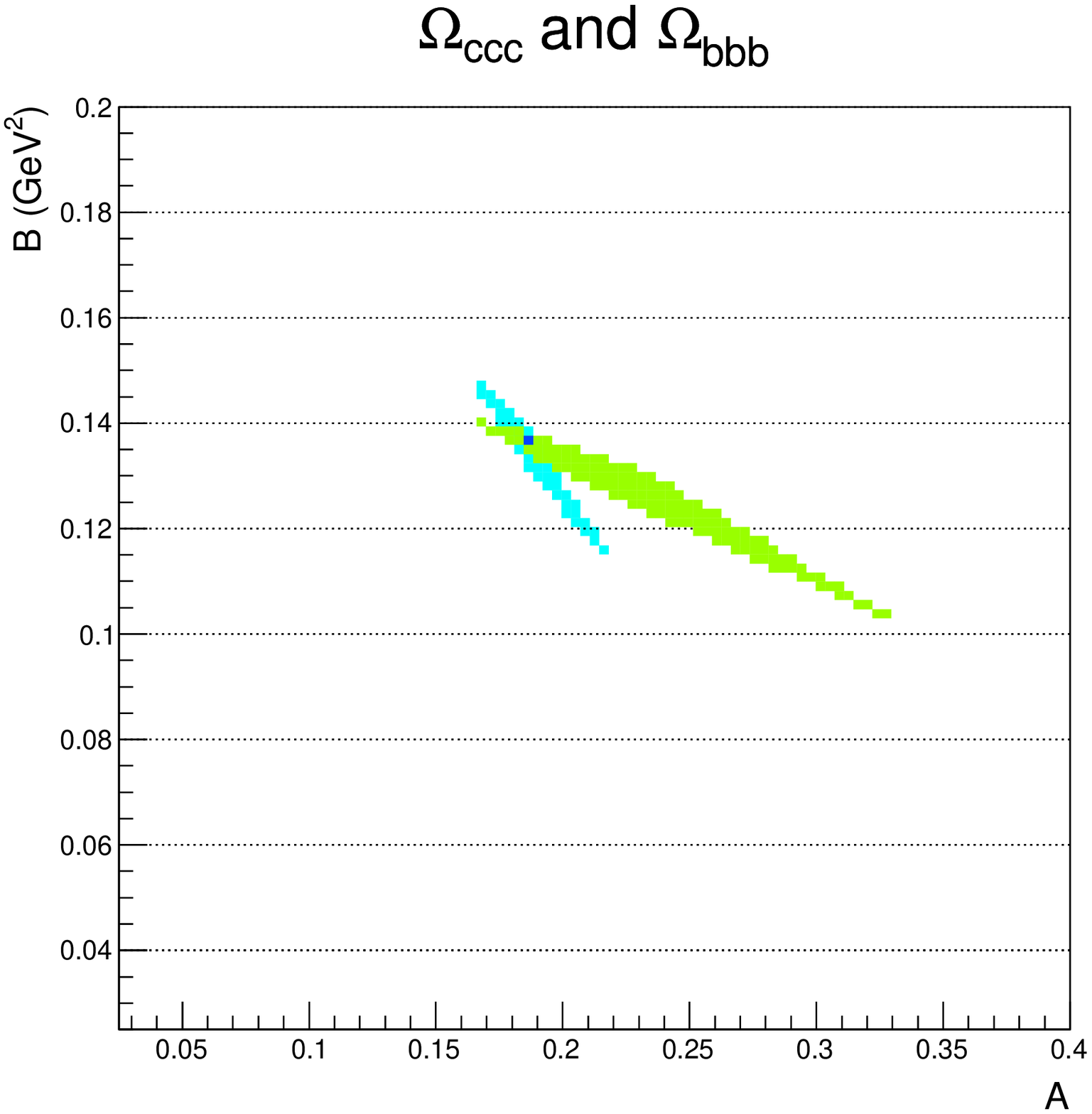}}
\caption{The light blue area stands for Coulomb and linear confining strengths reproducing the 
$E_2(3/2^+)$, $E_3(3/2^+)$, and $E_1(1/2^-)$ energies of the $bbb$ system.
The green surface represents the same for the $ccc$ system considering the spin-spin interaction
of Eq.~(\ref{Pot3QSS}). The dark blue point corresponds to our best
fit.}
\label{fig6}
\end{figure}

One may finally wonder if the Coulomb and linear confining 
strengths of the $3Q$ potential that we have found reproducing the 
nonperturbative lattice QCD results of Refs.~\cite{Mei12,Pad13} are 
unique. To answer this question we have considered the most relevant 
states, the first, $E_2(3/2^+)$, and second, $E_3(3/2^+)$, radial 
excitations and the first orbital excitation, $E_1(1/2^-)$\footnote{Having in mind the 
level degeneracies existing in the absence of rotational symmetry breaking, these
states represent the majority of the spectra~\cite{Mei12}.}. We have 
represented in Fig.~\ref{fig4} the sets of parameters that are able 
to simultaneously reproduce the positive and negative parity excited 
states for the $bbb$ system with a reasonable bottom quark mass, $m_b=4.655$ GeV.
As one can see there is a small range of parameters reproducing the 
general pattern of the $bbb$ spectrum. 
The strengths of the Coulomb and linear confining potentials we have 
determined giving rise to the results shown in Fig.~\ref{fig2} ($A=0.1875$ and $B=0.1374$ GeV$^2$),
are in the middle of the aforementioned area where the most important states are properly reproduced. 
In Fig.~\ref{fig5} we have represented the parameters describing the $bbb$ and 
the $ccc$ systems with a simple Coulomb plus a linear confining potential,
using a reasonable charm quark mass, $m_c=1.6$ GeV. There are mainly two noticeable aspects.
First of all, the smallness of the area reproducing the $ccc$ data. Even more, such area is 
in the limits of the uncertainties of the nonperturbative lattice QCD data for the $ccc$ system. 
The second relevant aspect is that there is no overlapping between 
the $ccc$ and $bbb$ range of parameters. As has been already noted,
these results could indicate that the spectrum of the $ccc$ system hardly 
accommodates to a simple linear confining plus Coulomb potential, needing of higher order terms in the interaction. 
If we repeat the same simulation for the $ccc$ system adding the spin-spin force of Eq.~(\ref{Pot3QSS}),
we get the results shown in Fig.~\ref{fig6}, where one can see the overlapping
between the $bbb$ and $ccc$ sets of parameters that coincides exactly in the Coulomb 
and linear strengths we have determined.

Let us finally comment on the possibility of applying the approach outlined in this work to 
light flavor sectors, as for example the excited spectrum of the $\Omega_{sss}$ baryon that has also been 
studied by means of lattice QCD~\cite{Bul10}. Unfortunately
the role played by other contributions beyond the confinement and the Coulomb 
potentials, as illustrated in Ref.~\cite{Val05}, leads to the 
conclusion that only baryons made of heavy flavors are clean 
enough to properly disentangle confinement and Coulomb contributions.

\section{Summary}
\label{Res}

To summarize, the spectra of baryons containing three identical heavy quarks, $b$ or $c$,
have been recently calculated in nonperturbative lattice QCD. 
The energy of the ground state and the lowest positive and negative parity excited states has 
been determined with high precision. These achievements constitute a unique opportunity to 
test phenomenological potential models in the regime in which they are expected 
to work best. We have analyzed these results by means of a Cornell-like potential 
using different numerical techniques for the three-body problem. 
For the case of the $bbb$ system a good agreement is obtained by means 
of a simple Coulomb plus linear confining potential. For the case of
the $ccc$ system the additional contribution of the spin-spin interaction is
needed to have a reliable mass for the charm quark. 
The spin-spin interaction comes suppressed by $M_Q^{-2}$, but it helps to correctly 
allocate the negative
parity excitations with respect to the radial excitations of the $3/2^+$ ground state.
As in the case of the heavy meson spectra, a larger value
of the Coulomb strength than predicted by SU(3) lattice QCD is needed.
The phenomenological strengths of the Coulomb potential reproducing the heavy 
meson and the triply-heavy baryon spectra satisfy $A/a<1/2$, slightly 
different from the 1/2 rule as the one-gluon exchange result. 
The strength of the linear confining interaction has to be also
larger than derived from SU(3) lattice QCD. 
Our results support a coherent
description of the $bbb$ and $ccc$ heavy-baryon spectra with the same Coulomb and 
confining strengths.

Let us conclude by emphasizing that the recent 
improvements in lattice QCD calculations of the heavy-baryon spectra~\cite{XXX15} may 
benefit of constituent quark model predictions, and, on the other hand, potential model approaches 
do also require guidance of lattice data for constraining not only the parameter space but 
also the possible functional forms to be explored.

\acknowledgments
We thank to S. Meinel for valuable discussion and 
information about the present status of nonperturbative lattice QCD calculations of excited heavy 
baryon states. We also thank to N. Mathur for providing us with 
the numerical data of the triply charm baryon spectrum.
This work has been partially funded by
Ministerio de Educaci\'on y Ciencia and EU FEDER under 
Contract No. FPA2013-47443 and by the
Spanish Consolider-Ingenio 2010 Program CPAN (CSD2007-00042) 
and by Generalitat Valenciana PrometeoII/2014/066. A.V. is thankful for financial 
support from the Programa Propio I of the University of Salamanca.

\end{document}